\documentclass[doublecol]{epl2} 
\usepackage{epsfig}

\title{Bound States and Universality in Layers of Cold Polar Molecules}
\shorttitle{Universal States in Layered Systems of Cold Polar Molecules}

\author{J.~R. Armstrong\inst{1} \and N.~T. Zinner\inst{1,2} \and D.~V. Fedorov\inst{1}, A.~S. Jensen\inst{1} }

\institute{                    
  \inst{1} Department of Physics and Astronomy - Aarhus University, Ny Munkegade, bygn. 1520, DK8000 \AA rhus C, Denmark\\
  \inst{2} The Niels Bohr Institute, Blegdamsvej 17, DK-2100 Copenhagen \O, Denmark \\

}

\abstract{The recent experimental realization of cold polar molecules in the rotational and vibrational ground state opens the 
door to the study of a wealth of phenomena involving long-range
interactions. By applying an optical lattice to a gas of cold polar
molecules one can create a layered system of planar traps. Due to the
long-range dipole-dipole interaction one expects a rich structure of
bound complexes in this geometry. We study the bilayer case and
determine the two-body bound state properties as a function of the interaction
strength. The results clearly show that a least one bound state will
always be present in the system. In addition, bound states at zero
energy show universal behavior and extend to very large radii. These
results suggest that non-trivial bound complexes of more than two
particles are likely in the bilayer and in more complicated chain structures in multi-layer systems.}

\begin{document}

\maketitle

\section{Introduction}
Quantum gases of polar atoms and molecules in their rovibrational
ground-state represent a unique opportunity to study the interplay of
long- and short-range interactions in the highly controllable trapped
gas environment. Early experiments used magnetic dipolar atoms
\cite{griesmeier2005,veng2008,fattori2008,pollack2009} which have
observable effects in spite of intrinsically weak dipole moments. Recently,
heteronuclear molecules with very large electric dipole moments have
been realized by a number of groups
\cite{ospelkaus2008,ni2008,deiglmayr2008,lang2008,ospelkaus2010}. The
goal of a quantum degenerate system of polar molecules with strong
$1/r^3$ long-range dipole-dipole forces therefore seems close at hand.

The attractive force of polar molecules in the head-to-tail
configuration can lead to collapse of the system
\cite{lushnikov2002}. However, as suggested by Wang et
al. \cite{wang2006}, a one-dimensional optical lattice that creates a
multilayered stack of pancake systems can stabilize the situation. If
we apply a field to polarize the dipoles perpendicular to the layers
then the intralayer interaction will be purely repulsive, whereas the
interlayer part will be attractive but with the optical lattices
separating the dipoles in different layers. This setup is presently 
being implemented experimentally.  As discussed in
\cite{wang2006}, the dipole-dipole force forms bound chains
and the system effectively behaves as a liquid of chains with
resemblance to rheological fluids. In the case of bosons we expect the
chains to Bose condense at low temperatures. However, if we have
fermionic polar molecules the situation is less clear since one would
expect a Bose-Fermi mixtures with various bound complexes
\cite{santos2010}.  The dipole potential can also now be inverted to have a repulsive core with the use of laser fields \cite{cooper2009,sarma2009}, which offers different physics possibilities than the ``natural'' dipole orientation.

If we simplify the problem to consider just two adjacent layers we have a system
with $1/r^3$ interactions that mimics the long-range 
$1/r$ interactions in graphene
\cite{novoselov2006} and semiconductor bilayers \cite{ye2010}. In the
semiconductor case bound states of exciton pairs with non-zero dipole
moments have been considered in connection with organic interfaces and
quantum wells \cite{yudson1997}. For small coupling strength it was
concluded that no bound state exists \cite{yudson1997}. This was also
stated in several recent works concerning cold polar molecules
\cite{wang2006,wang2007,santos2010} where the conclusion was based on a Gaussian
ansatz. However, at small coupling the particles are strongly
delocalized and a localized Gaussian is therefore not a good
approximation \cite{jen04}. To make matters worse, the potential
integrates to zero over the plane and thus 
at small coupling the Landau criterion \cite{landau1977} for
a bound state in two-dimensional systems is not applicable.  
Using scattering theory it was recently shown that
a bound state presumably exists for arbitrarily small moments
\cite{wang2009,ticknor2009}.  However, the scattering theory is
intricate and does not yield straightforward information about the
behavior of the two-body bound state wave function.

The purpose of the present work is to compute and explain the basic
properties of two-body systems used as the fundamental building blocks
for layered dipolar structures.  We shall employ simple model
potentials to extract universal properties, point out where details of
the potentials are needed, and illustrate relations between wave
functions, energies, and radii.  We shall use square well, harmonic
oscillator and $1/r$-potentials in two dimensions, and compare to
solutions of the true dipole-dipole potential. We briefly sketch the
model solutions, discuss energies, threshold properties, and various
implications.

\begin{figure}
\centerline{\epsfig{file=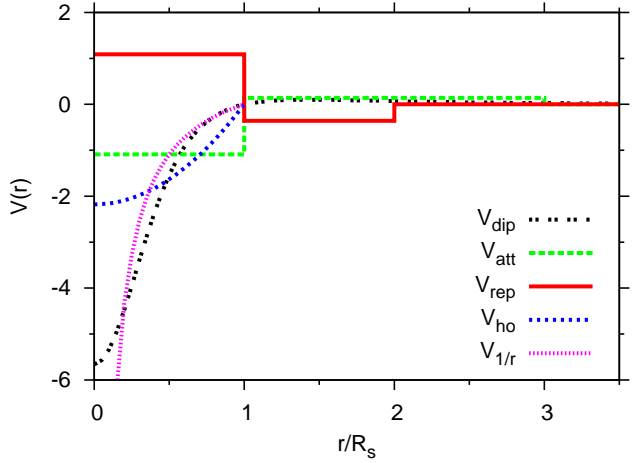,scale=1.2}}
\caption{Potentials of different shapes, dipole-dipole potential ($V_{dip}$), 
two square wells ($V_{att}$, $V_{rep}$), harmonic oscillator
($V_{ho}$), and $1/r$ ($V_{1/r})$. All potentials have the nodes at
$r=R_s$, the same attractive and repulsive volumes, and net volume
zero. }
\label{fig1}
\end{figure}

\section{Model solutions}

We solve the 2-dimensional (2D) Schr\"odinger equation, which is
possible to do analytically with several model potentials.  We use
cylindrical coordinates $(r,\theta)$ and separate the total
wave function $\Psi=R(r)\Phi(\theta)$ into radial $R(r)$ and angular
$\Phi(\theta)$ parts.  With $\Phi(\theta) =
\exp(im\theta)/\sqrt{2\pi}$, where $m$ is an integer, the stationary
radial Schr\"odinger equation becomes:
\begin{equation}
\left[-\frac{\hbar^2}{2m}\left(\frac{d^2}{dr^2}+\frac{1-4m^2}{4r^2}\right)
 +V(r)\right]u(r)=Eu(r),
\label{schrod1}
\end{equation}
where $u=\sqrt{r}R(r)$ is the reduced radial wave function, $E$ is the
energy, and $V(r)$ is the potential, assumed to be spherically symmetric. We consider
potentials with net volume equal to zero, $\int V(r)d^2r=0$, just as for
the dipole-dipole potential, i.e. 
\begin{equation} \label{e25}
 V(r) = D^2\frac{r^2-2d^2}{(r^2 + d^2)^{5/2}} \; ,
\end{equation}
where $D$ is the electric dipole moment and $d$ is the distance
between the two different layers containing the particles.  This
potential is in Fig.~\ref{fig1} compared with square well, harmonic
oscillator, and $1/r$ potentials at distances less than their common
node at $R_s \equiv d\sqrt{2}$.  The equal volume conditions relate
strengths and shifts for given radial shapes. Thus we have $D^2=V_s
R_{s}^{3}(3/2)^{3/2}/2$, and for the square well shape, the small,
$V_s$, and large, $V_l$, distance absolute strength values are related
by
\begin{equation}
\frac{V_s}{V_l}=\left(\frac{R_l}{R_s}\right)^2-1 \;,
\label{PotVol}
\end{equation}
where $R_l$ is the radius where the outer square well ends.  The attractive
part is the most interesting and we shall only use the harmonic and
$1/r$ potentials with the square barrier shape for $R_l>r>R_s$ as illustrated in Fig. \ref{fig1}.

One of our main concerns is the appearance of a bound state at small
couplings. We therefore only consider the most attractive $m=0$
potential which simplifies to
\begin{equation}
\frac{d^2u}{dr^2}+\left(\frac{1}{4r^2}+k^2\right)u=0,
\label{schrod2}
\end{equation}       
where $k^2=2m(E-V(r))/\hbar^2$ depends on $r$.  

Let us now first solve completely the piecewise constant potential in 
Fig.~\ref{fig1}
for use as
a reference standard.  Then the wave functions in the three different
regions of space are the Riccati Bessel functions of order $-1/2$,
which means that the solutions for $R(r)$ are various Bessel functions
of order 0 depending on the region of space:
\begin{equation}
R=\left\{ \begin{array}{ll}
AJ_0(k_1r) & r\leq R_s \\
BI_0(k_2r)+CK_0(k_2r) & R_s< r \leq R_l \\
DK_0(k_3r) & r>R_l\;,
\end{array}\right.
\label{WFs}
\end{equation}
where $J_0,I_0,K_0$ are ordinary and modified Bessel functions of the
first and second kind, respectively.  The coefficients $A, B, C,$ and
$D$ are determined by matching at the region boundaries and by
normalization. The wavenumbers, $k_i$, are the absolute values of $\bm
k$ in the regions.  Matching logarithmic derivatives leads to the
transcendental equation for the energies:
\begin{eqnarray}
\frac{k_2R_l K_0(k_3R_l) K_1(k_2R_l)-k_3R_l K_0(k_2R_l) K_1(k_3R_l)}
{k_3R_l I_0(k_2R_l)K_1(k_3R_l)+k_2R_l K_0(k_3R_l)I_1(k_2R_l)}=\nonumber \\
\frac{k_2R_s J_0(k_1R_s)K_1(k_2R_s)-k_1R_s K_0(k_2R_s)J_1(k_1R_s)}
{k_2R_s J_0(k_1R_s)I_1(k_2R_s)+k_1R_s I_0(k_2R_s)J_1(k_1R_s)}.
\label{e35}
\end{eqnarray}
This formula and the wave function, Eq.(\ref{WFs}),
are valid for the potential $V_{att}$ in Fig. \ref{fig1}.  
For $V_{rep}$ from Fig. \ref{fig1}, one merely
takes the analytic continuation of the relevant Bessel function.  
Expanding the solution for small strengths results in 
\begin{equation} 
-E=E_0\exp\left[-\frac{E_0}{V_l}\left(2+\frac{E_0}{V_l\ln(R_l/R_s)}\right)\right],
\label{expansion}
\end{equation}
where $E_0=2\hbar/(mR_l^2)$. Eq.(\ref{expansion}) with the second
order potential strengths replaces the Landau expression when the
potential integrates to zero\cite{landau1977}.

Analytic solutions can also be found when the short-distance part of
the potential is substituted by a harmonic oscillator potential,
$V_{ho}(r)=2V_s[(r/R_s)^2-1]$ and a $1/r$ potential,
$V_{1/r}(r)=V_s(1-R_s/r)$. In both cases the potentials are zero for
$r=R_s$ and $V_s$ is the strength of the box potential with the same
volume below $R_s$. The radial wave functions for $r<R_s$ are given by
$R(r) = N \exp(-z/2) M$, where $M$ is the confluent hypergeometric
function, or Kummer function $M(a,b,z)$ with three arguments. We have
$b=1$ and $a$ and $z$ are:
\begin{eqnarray}
a&=&\frac{1}{2}-\frac{E+2V_s}{4\hbar}\sqrt{\frac{mR_s^2}{V_s}}\;,\;\; 
 z= \frac{2r^2\sqrt{mV_s}}{\hbar R_s} \\
a&=&\frac{1}{2}-\frac{R_sV_s}{\hbar}\sqrt{\frac{m}{2(V_s-E)}}\;,\;
 z= \frac{r\sqrt{8m(V_s-E)}}{\sqrt{\hbar^2}}  \;
\end{eqnarray}
for oscillator and $1/r$ potentials, respectively. This is the
solution vanishing at $r=0$. Matching at $r=R_s$ with the square
well solutions for $r>R_s$ leads to trancendental equations for the
energies.

\begin{figure}
\centerline{\epsfig{file=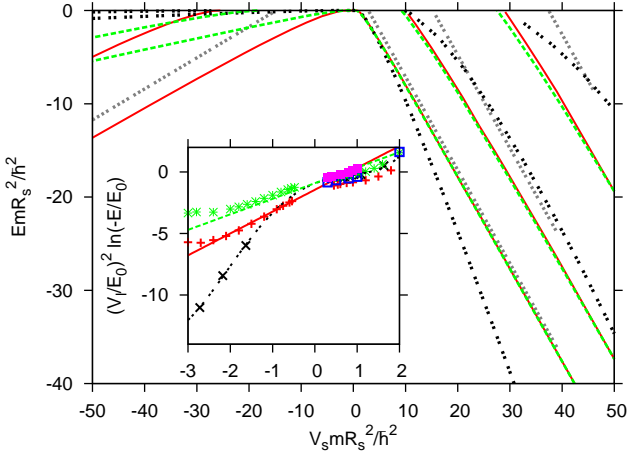,scale=1}}
\caption{Energies of states as a function of $V_s$ in 
units of $\hbar^2/(m R_s^2)$.  Right and left correspond to $V_{att}$
and $V_{rep}$, respectively, of Fig. \ref{fig1}.  The dotted grey
lines are the calculated asymptotic values for the square well states.
The solid red lines are the square well energies with $R_l=2R_s$, the
dashed green lines are square well energies with $R_l=3R_s$, and the
black double-dotted line the dipole-dipole potential results. The
inset is a zoom of small energy plotted logarithmically (values in
points colored as in the main plot), the lines are from
Eq. (\ref{expansion}). The magenta and blue points, from $1/r$ and
oscillator potentials, appear together with the dipole points found for
$R_l=3R_s$($V_l$ is from Eq.~(\ref{PotVol})), and the green square well points.  The black line 
for the dipole potential is to guide the eye.}
\label{EV1fig}
\end{figure}

\section{Energies}

The energies obtained from Eq. (\ref{e35}) are shown in
Fig.~\ref{EV1fig} as function of $V_sR_s^2$, or equivalently of volume
of either attraction or repulsion.  We choose $R_l=2R_s$ or $R_l=3R_s$, as indicated, and use $V_l$
from Eq.(\ref{PotVol}).  Increasing $V_s$ for both natural and inverted
potentials leads to more bound states which more and more are
determined from the attraction alone and independent of the confining
barriers.  The condition, $J_0(k_1R_s)=0$, determines $k_1$ for
$V_{att}$ in Fig. \ref{fig1} and thereby the bound state energy $E_n$
is related to the nodes of the Bessel function, i.e.
\begin{equation}
E_n=-V_s+\frac{\hbar^2(j_{0,n})^2}{2mR_s^2},
\label{besselE}
\end{equation} 
where $j_{0,n}$ is the $n$th zero of $J_0(x)$.  This asymptotic limit
of straight lines of slope equal to one is valid for an infinitely
high barrier.  The ground state is the deepest and agrees well with
the asymptotic limit, but for the excited states the well must be
deeper to reach the limit, around -150 for 1\% agreement for the first
excited state. 

For $V_{rep}$ in Fig. \ref{fig1}, where the repulsion is for $r< R_s$
and the attractive well is at $R_s<r< R_l$, no confining barrier
exists at larger distance.  Here, the bound state wave function is
$R(r)=AJ_0(r)+BN_0(r),$ where $N_0(r)$ is the Neumann function.  In
the deep well limit, the wave function must vanish at both endpoints
where large energy implies large arguments of both the Bessel and
Neumann functions.  Then $k_2^2(R_l-R_s)^2=n^2\pi^2$ or equivalently
\begin{equation} \label{e75}
E_n=-\frac{V_s}{(R_l/R_s)^2-1}+\frac{n^2\pi^2\hbar^2}{2m(R_l-R_s)^2},
\end{equation}
where the slopes of the lines are dependent on $R_l$ and therefore different 
from the $V_{att}$ case. These estimates are much further off than the similar estimates for $V_{att}$.
For the difference between the energies and the estimate to be 1\% for
the ground state, the depth of the well needs to be around 100.  For
that level of agreement in the first excited state, the well depth
needs to be around 250.

The result for small strengths are shown in the inset of
Fig.\ref{EV1fig}.  One bound state is always present for all
potentials even when the strength is approaching zero. The dependence
on both strengths and radius ($R_l$) is substantially stronger for the
inverted potentials.  The very small energies close to threshold obey
the limiting linear dependence from Eq. (\ref{expansion}).  They are
extremely small due to the zero net volume, and hence difficult to
obtain accurately.  The numerical results for the dipole potential almost coincide with
the energies for the other potentials. On the scale of
the inset the energies are exceedingly small and difficult to
calculate numerically, especially for the inverted dipole potential.  For larger strengths $mV_s R_s^2/\hbar^2 >2$ on the inverted side and
$mV_s R_s^2/\hbar^2 > 10$ on the natural side, we found that the
energies rather precisely are given by $E=-c_0 D^2/d^3\exp(-c_1\hbar^2
d/(mD^2)$, where $(c_0,c_1)=(0.85,9.86),(0.012,19.9)$ for the natural
and inverted potentials, respectively.

The values in the experiments \cite{ni2008} 
correspond to strengths of about $V_s R_{s}^{2}m/\hbar^2=1.9$ or
$m R_{s}^{2} E/\hbar^2\sim -0.6$ (Fig.~\ref{EV1fig})
which is on the verge of universality (see Fig. \ref{Er2}), although
current temperatures are too high to maintain 
such a weakly bound state \cite{ni2008}. Fortunately, systems with 
larger $D^2/d^3$ can be explored where
the bound state energy is much larger and obeys the analytic formula given
above.

\begin{figure}
\centerline{\epsfig{file=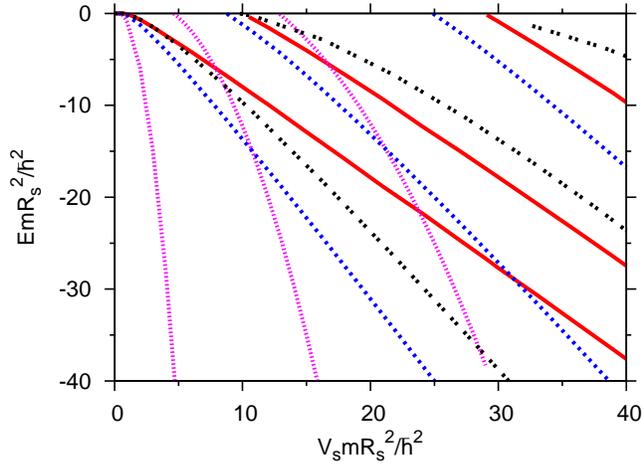,scale=1.2}}
\caption{The lowest three bound state energies as function of strength
for different potentials, square well (solid red), harmonic oscillator
(dashed blue), $1/r$ (dotted magenta), and dipole (double-dotted black).
The square well barrier is the same in all cases, $R_l=2R_s$.}
\label{diffPot}
\end{figure}

These computations are also carried out for harmonic oscillator and
$1/r$ potentials at small distance with a square well at larger
distances.  The condition of exponentially vanishing wavefunction with
$r$ implies that the first argument $a$ of the Kummer function must be
a non-positive integer $-n$ counting the bound states. Then
$M$ reduces to the Laguerre polynomials.  The energies become
\begin{eqnarray}
E_{ho}&=&2[-V_s+\sqrt{\frac{V_s\hbar^2}{mR_s^2}}(2n+1)], \label{e45} \\
E_{1/r}&=&V_s-\frac{2mV_s^2R_s^2}{\hbar^2(2n+1)^2}, \label{e55}
\end{eqnarray}
where $n$ can take the values 0,1,2, etc. These energies are only
approached asymptotically as for the square well in Fig.~\ref{fig1}.
They are highly potential dependent for well-bound states as the
barrier essentially has no influence on these energies, see
Fig.~\ref{diffPot}.  The potential in Eq.(\ref{e25}) leads to energies
between those of harmonic and square well potentials. Thus limits to
realistic potentials can be provided by analytic models.

However, in the limit where $V_s\rightarrow 0$ both energies in
Eqs.(\ref{e45}) and (\ref{e55}) approach zero from positive values. The
approximations are too crude, and we replace the $r>R_s$ potentials by the
corresponding square well.  We then find numerically that the energies
for both these potentials always remain negative corresponding to bound
states for all values of $V_s$. Thus all the investigated potentials 
with zero net volume always have at least one bound state.

\section{Threshold properties}

The weakly bound states often reveal unique physics as for example
Efimov and halo states with universal properties.  The number of nodes
as function of volume can be found rather precisely from
Eqs.(\ref{besselE}-\ref{e55}) by solving for $n$
with $E_n=0$. The explicit result for harmonic oscillator, $2n+1
= \sqrt{m|V_s|R_s^2}/\hbar$, whereas for the $1/r$
potential the right-hand side is a factor $\sqrt{2}$ larger. 
Thus these potentials would have the same number of bound
states if the volume of the oscillator were twice as large as that of
the $1/r$-potential.  The dipole-dipole potential seems from
Fig.~\ref{fig1} to have intermediate properties. However, the 
dependencies on strength for each energy differs substantially, see
Eqs.(\ref{e45}) and (\ref{e55}).

The same extrapolation to the threshold can be applied to the
wave functions. We then find that the mean square radius is given as
$<r^2/R_s^2> = 1/2$ for the oscillator and $5/8$ to leading order for
the $1/r$ potential.  These threshold radii, obtained from asymptotic
strong binding, are however qualitatively completely wrong when the
energy is sufficiently close to zero.  In Fig.~\ref{EV1fig} the
straight lines approaching zero bend over as the system attempts to
stay bound for a weaker potential.  The wave function is
correspondingly leaking out under the barrier as the energy approaches
zero.  This is the effect producing nuclear halos
\cite{jen04} where the mean square radius for two particles in three 
dimensions becomes inversely propertional to the energy. The analogue
here is that most of the probability is found for $r>R_l$ where the
wave function is $K_0$. This means that the mean square radius then
approaches \cite{nie01}
\begin{equation} \label{e85}
  <r^2> =  \frac{\hbar^2}{2m|E|} \frac{\int_0^{\infty} x^3 |K_0(x)|^2 dx }
 {\int_0^{\infty} x |K_0(x)|^2 dx }   = \frac{1}{3} \frac{\hbar^2}{m|E|} \;,  
\end{equation} 
which is a universal result independent of both the particular ($s$-wave)
state considered and the shape of the attractive potential.

\begin{figure}
\centerline{\epsfig{file=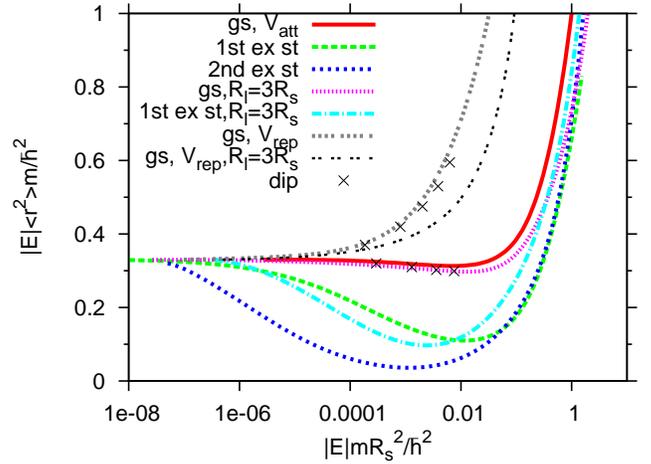,scale=1.2}}
\caption{Binding energy multiplied with $<r^2>$ plotted against the 
binding energy for the lowest bound states for the square well
(parameters in the legend).  The crosses are results for the dipole
potential.  The scale on the abscissa is logarithmic, with small
binding energies to the left.}
\label{Er2}
\end{figure}

The rate of approach to the asymptotic value in Eq.(\ref{e85}) is seen
in Figure \ref{Er2}.  Large binding energies correspond to wave
functions located in the attractive well. As the threshold is
approached the wave function begins to leak out and eventually ends up
in the universal limit in Eq.(\ref{e85}).  However, how that limit is
approached depends on the presence of a barrier. Without a barrier
($V_{rep}$ from Fig. \ref{fig1}), all states, both ground and excited,
approach the weak binding limit in the same way.  This is precisely
the halo effect \cite{jen04}.

The similar approach to universality is found for the ground state
for  $V_{att}$ from Fig. \ref{fig1} since then the barrier vanishes with the binding energy.
For excited states of $V_{att}$, the barrier remains finite for
vanishing binding energy.  High excitation goes together with high
barrier which causes even the weakly bound states to remain localized
in the attractive region.  The approach to universality is delayed by
many orders of magnitude in binding energy.  During the approach the
radius remains small for small energies and consequently the curves
dive below the universal limit before the eventual approach.  These
features become more pronounced with excitation energy. 

In Figure \ref{Er2} we also show results for different values of
$R_l$, still with $V_l$ adjusted to maintain zero net volume.  The
curves for the ground states are roughly identical. This is also seen
at higher energies for the first excited state, but eventually at
small energies a wider barrier delays the escape of the wave function
even though the barrier is correspondingly smaller.

\begin{figure}
\centerline{\epsfig{file=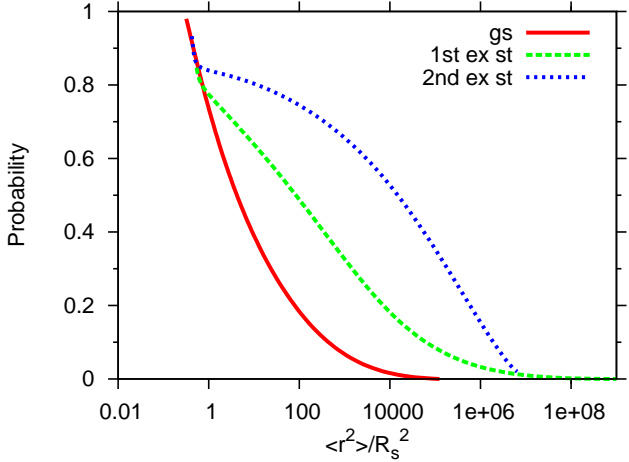,scale=1.20}}
\caption{ The probability, $\int_0^{R_s}|u|^2dr$, of being in the 
attractive part of the $V_{att}$ potential for the lowest three bound
states shown as a function of $\langle r^2 \rangle$.  For $V_{rep}$,
all states behave as the ground states of both potentials.}
\label{R2}
\end{figure}

The rate of approach to the universal limit at threshold is further
illustrated in Figure \ref{R2}.  The probability distribution from the
wavefunction is moving slower than the distribution in the mean square
radius integral.  The bulk of the probability can stay in the
attractive well while the root mean square radius is much larger than
the radius of the attraction.  Thus the tail properties are crucial.
These effects are enhanced for excited states because the final
expulsion of the wave function to the external region comes much later
for states constraint by a large barrier.

The formal connection is that the wave function essentially is $J_0$
until the energy is very small where it has to change to $K_0$, see
Eq.(\ref{WFs}). Computation of mean square radius then employs $J_0$
for normalization but $K_0$ for the $r^2$-distribution. Thus
neither short nor large-distance properties are sufficient for a
description in this transition region.

\section{Implications for many particles}

Several conclusions are immediately deduced from our model systems.
First of all, at least one bound state must appear in 2D for any
potential with zero net volume.
Thus, in contrast to the statements
in \cite{yudson1997,wang2006,wang2007,santos2010}, one bound state is
always present in a bilayer with dipoles oriented perpendicular to the
layers.  This in turn means that arguments based on the existence of a
critical strength for binding should be re-considered.  Furthermore,
since a large dipole moment implies a large strength, several
two-body bound states will be present. This must be taken into account
in simulations of configurations of actual systems where a finite
temperature might lead to population of different excited states.
In practice, finite temperature in the system will put a natural limit
on how small binding energies one needs to consider, i.e. for
$|E|<k_B T$ the bound states are thermally dissociated into the continuum 
and therefore largely irrelevant.

Second, a positive net volume still allows bound states in two
dimensions.  However, now it is necessary to have a non-zero minimum
attraction where the strength would increase with the net volume.
This resembles the situation in three-dimensional quantum mechanics.

Third, the structure of many particles in layers are strongly
influenced by their pair interactions.  The absence of a critical
interaction strength for binding implies that a phase transition from a superfluid to a dipolar chain liquid cannot
occur in contrast to the suggestion in \cite{wang2006,wang2007}.
Furthermore, the presence of bound states in bilayers for all
strengths immediately implies that chains in multilayers will also be
present for all couplings.  We would therefore expect the system
to always be a dipolar chain liquid.  As all the individual chains can
form immediately the chain-chain interaction becomes interesting and
necesary to include in careful future investigations. 

We can in fact give an upper bound on the binding energy of a chain of $M$ 
particles in $M$ different layers. The scaling of two-body energy 
with layer distance is $E(nd)=E_2/n^2$, with $n$ an integer giving 
the distance in equally spaced layers a distance $d$ apart. Here $E_2$ 
is the two-body energy we have calculated above which depends on the 
coupling strength. Taking the large $M$ limit, the upper bound is 
\begin{equation}
E_M=E_2\,M\sum_{n=1}^{\infty} \frac{1}{n^2}=\frac{\pi^2}{6}M\,E_2.
\end{equation}
This estimate of the energy agrees well with the calculation of \cite{wang2006} where the harmonic
approximation was at small couplings for $M=2$ and $M=81$.

Fourth, the inverted potential with $V_{rep}$ in Fig. \ref{fig1} 
with repulsion at small distances and an intermediate attractive pocket
also has at least one bound state for any strength.  In the 
bilayer system with a finite density of fermionic
polar molecules in each layer, Ref.~\cite{sarma2009} 
found an interesting particle-hole
coherent state.  The density used there was chosen small so 
intralayer interaction is negligible and the focus was on the 
interlayer repulsion corresponding to a barrier at small radii and
an attractive pocket outside. In \cite{sarma2009} the presence of 
bound states shown in this paper was ignored.
In a real system at finite temperature this coherent state could
probably exist at low dipole strength. However, at larger strengths
the bound states must be taken into account and one would presumably
have instead an interacting gas of bound pairs behaving as a
superfluid.

Fifth, the bound states become extremely extended as the energy gets
sufficiently close to zero. Such delocalized two-body
states can in turn enter into complicated multi-particle complexes\cite{wun10}.  A
bound three-body system of two particles in one plane and one in
between in the adjacent plane is therefore possible. This presupposes
that the repulsive intraplane interaction between these extended
structures is less than two times the two-body binding energy that
created the two-body bound states.  This sort of Y-junction
configuration can be very interesting in thermodynamic considerations
of chains as it will contribute non-trivially to the entropy and can help lower
the free energy.  Full quantum studies of such configurations are
therefore very relevant and worth pursuing.

Sixth, the optimum conditions for interesting multi-particle
structures are probably in the regime close to the threshold for
binding of the second state. This can be below the threshold where
additional attraction from other particles would lead to binding in
analogy to Borromean three-body systems where the two-body subsystems
are unbound \cite{nie01,jen04}.  It can also be for slightly larger
attraction and with a bound two-body state since such a system still is
spatially extended and in a sense rather similar to the unbound
continuum state. The latter case is analogous to the extremely weakly
bound atomic helium dimer where the trimer becomes well bound but with
a very weakly bound and spatially extended excited state
\cite{nie01,jen04}.

Seventh, the regime of weak binding, strong delocalization and large
root-mean-square radius exhibits universal features independent of
the shape of the potentials. The same type of universality is likely
to exist for multiple bound states but much finer tuning is probably
required to reach these structures.

\section{Conclusions}

Presently experimentalists are working to produce layered systems of dipolar molecules.  
We use simple model potentials to study the bilayer case with
dipoles polarized perpendicular to the layers.  We find the solutions and calculate properties of the wave functions and in
particular binding energies and radii as functions of dipole moment or
strength of the potentials. Realistic potentials are used to test the
generality of our results.

We conclude that there always is a bound state for all strengths of
the dipole-dipole potential.  We find that the wave functions of both
ground and excited states show universal behavior at zero energy as
they basically reside where the potential has become vanishingly
small.  To access this universal regime the dipole strength must be
tuned around the threshold for a bound state to appear.  The extended
wave functions indicate that three or more particle complexes are
possible in chains and in bilayers.  The repulsive in-plane
interaction becomes interesting in connection with these structures. 
In any case the tuning of interactions to universal regimes emphasizes
the close analogy to the physics studied through the well-known
technique of Feshbach resonances.

\acknowledgments

NTZ is grateful to B. Wunsch, D.-W. Wang and E. Demler for numerous discussions.

\end{document}